\documentclass[prl,twocolumn,showpacs,superscriptaddress]{revtex4-1}

\usepackage[normalem]{ulem}
\usepackage{graphicx}
\usepackage{amssymb,amsmath,amstext,latexsym,bm}
\usepackage[usenames,dvipsnames]{xcolor}
\usepackage[colorlinks=true,citecolor=Blue,linkcolor=RubineRed,urlcolor=Blue]{hyperref}

\newcommand{\phdagger}[0]{{\phantom{\dagger}}}

\newcommand{\ket}[1]{| #1 \rangle}

\newcommand{\av}[1]{\langle #1 \rangle}

\newcommand{\Av}[1]{\left \langle #1 \right \rangle}

\def\rmi{{\mathrm{i}}}

\DeclareMathOperator{\Tr}{Tr}

\begin{document}

\title{Large Deviations Beyond the Kibble-Zurek Mechanism}
\author{Federico Balducci\href{https://orcid.org/0000-0002-4798-6386}{\includegraphics[scale=0.05]{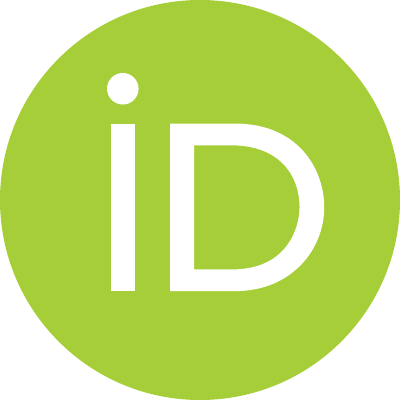}}}
\affiliation{Department of Physics and Materials Science, University of Luxembourg, L-1511 Luxembourg, G.\ D.\ Luxembourg}
\author{Mathieu Beau\href{https://orcid.org/0000-0003-0502-6210}{\includegraphics[scale=0.05]{orcidid.pdf}}}
\affiliation{Department of Physics, University of Massachusetts, Boston, MA 02125, USA}
\author{Jing Yang\href{https://orcid.org/0000-0002-3588-0832}{\includegraphics[scale=0.05]{orcidid.pdf}}}
\affiliation{Department of Physics and Materials Science, University of Luxembourg, L-1511 Luxembourg, G.\ D.\ Luxembourg}
\affiliation{Nordita, KTH Royal Institute of Technology and Stockholm University, Hannes Alfv\' ens v\"ag 12, SE-106 91 Stockholm, Sweden}
\author{Andrea Gambassi\href{https://orcid.org/0000-0003-3450-6125}{\includegraphics[scale=0.05]{orcidid.pdf}}}
\affiliation{SISSA -- International School for Advanced Studies, via Bonomea 265, 34136 Trieste, Italy}
\affiliation{INFN, Sezione di Trieste, Trieste, Italy}
\author{Adolfo del Campo\href{https://orcid.org/0000-0003-2219-2851}{\includegraphics[scale=0.05]{orcidid.pdf}}}
\email{adolfo.delcampo@uni.lu}
\affiliation{Department of Physics and Materials Science, University of Luxembourg, L-1511 Luxembourg, G.\ D.\ Luxembourg}
\affiliation{Donostia International Physics Center, E-20018 San Sebasti\'an, Spain}

\begin{abstract}
   The Kibble-Zurek mechanism (KZM) predicts that the average number of topological defects generated upon crossing a continuous or quantum phase transition obeys a universal scaling law with the quench time. Fluctuations in the defect number near equilibrium are approximately of  Gaussian form, in agreement with the central limit theorem. Using large deviations theory, we characterize the universality of fluctuations beyond the KZM and report the exact form of the rate function in the transverse-field quantum Ising model. In addition, we characterize the scaling of large deviations in an arbitrary continuous phase transition, building on recent evidence establishing the universality of the defect number distribution.
\end{abstract}

\maketitle

The Kibble-Zurek mechanism (KZM) is an important paradigm in nonequilibrium statistical physics, describing the dynamics across a continuous phase transition~\cite{Dziarmaga10,Polkovnikov11,DZ14}. The divergence of the equilibrium relaxation time in the neighborhood of the critical point makes the critical dynamics necessarily nonadiabatic for large systems and leads to the spontaneous formation of topological defects. Consider a phase transition from a high symmetry phase to a broken symmetry phase, induced by varying a control parameter $g$ across its critical value $g_c$ in a finite quench time $\tau_Q$. The central prediction of the KZM is that the average defect density, generated during the phase transition, displays a universal power-law dependence as a function of the quench time. The KZM thus makes a quantitative prediction on the breakdown of adiabatic dynamics across a phase transition and holds both in the classical and quantum regimes~\cite{Polkovnikov05,ZDZ05,Damski05,Dziarmaga05,ZD08,Dziarmaga10,Polkovnikov11,DZ14}.

Kibble's pioneering work was motivated by cosmological considerations regarding structure formation in the early universe~\cite{Kibble76a}. The prospect of exploring analogous phenomena in condensed matter systems was soon realized~\cite{Kibble76b,Zurek85,Zurek96c} and pursued experimentally~\cite{DZ14,Keim15,Maegochi22,Du23}. The advance of quantum technologies has led to new tests of the KZM using quantum simulators in a variety of platforms, including ultracold gases~\cite{Weiler08,Lamporesi13,Navon15,Anquez16,Ko19,ShuaiChen20,Duan20}, trapped ions~\cite{Ulm13,Pyka13,Cui16,Cui20,Duan23}, and Rydberg gases~\cite{Keesling19,Ebadi21}. Recently, the KZM has been studied with quantum computing devices, such as quantum annealers~\cite{Gardas18,Weinberg20,Bando20,King22}. The accumulated body of literature broadly supports the validity of KZM in a wide variety of systems. 

Experiments probing critical dynamics generally involve an ensemble of single experimental runs or individual realizations in which measurements are performed. As a result, they can access information beyond the average defect density and characterize the ensemble statistics. It is thus natural to ask whether there are universal signatures in the statistical properties of spontaneously generated topological defects~\cite{Cincio07,delCampo2018Universal,delcampo22}. The full counting statistics of defects appears to be universal in classical and quantum systems. Specifically, in classical continuous phase transitions, it has been found that the defect number distribution is binomial with an average density in agreement with the KZM~\cite{GomezRuiz20,Mayo21,delcampo21,GomezRuiz22}. Exact solutions in quantum integrable systems have shown that the kink number distribution is Poisson-binomial~\cite{delCampo2018Universal,Cui20}, a feature that can hold even when the system is coupled to an environment~\cite{Bando20,King22}. These predictions build on the conventional KZM but lie outside its scope, requiring additional assumptions. We shall thus refer to them as beyond-KZM physics. 

The average number of defects is an extensive quantity. By contrast, the defect density is intensive, and its fluctuations near equilibrium are approximately Gaussian, in agreement with the central limit theorem. Large deviations theory (LDT) addresses the probability of nontypical events in which an intensive quantity deviates from its average value. The probability of such large deviations decays exponentially with increasing system size, at a rate controlled by the so-called rate function~\cite{Ellis06,Touchette09,Dorlas21}. LDT provides a building block of statistical mechanics in and out of equilibrium. As such, it is a natural framework to explore beyond-KZM physics. To date, LDT has been used to describe the dynamics of many-body quantum systems in the limit of sudden quenches when $\tau_Q\rightarrow 0$, e.g., to characterize the work statistics of a given process~\cite{Gambassi12,Goold18,Perfetto19}.

In this Letter, we establish the universality of large deviations beyond the KZM, after crossing  a quantum phase transition in a finite time. Specifically, we report the exact rate function of the driven transverse-field quantum Ising model (TFQIM), characterizing the statistics of large fluctuations away from the mean kink density predicted by the conventional KZM. We further generalize these results  to characterize the universality of large deviations in an arbitrary continuous phase transition leading to point-like defects.


{\it The transverse-field quantum Ising model.} 
The TFQIM has been instrumental in generalizing the KZM from the classical to the quantum domain~\cite{Polkovnikov05,ZDZ05,Damski05,Dziarmaga05,Cui16}, and assessing the universality of beyond-KZM physics, both in theory~\cite{delCampo2018Universal} and experiments~\cite{Cui20,Bando20,King22}. Its Hamiltonian is given by 
\begin{equation}
    H[g(t)]=-J \sum_{l=1}^N \left[g(t)\sigma_l^x + \sigma_l^z \sigma_{l+1}^z\right]\label{eq:Ising},
\end{equation}
where $J>0$ favors ferromagnetic alignment and $g(t)$ plays the role of an effective magnetic field. In the fermionic representation, the Ising chain Hamiltonian becomes~\cite{SupplInf}
\begin{equation}
    \label{eq:H_TFIM_fermions}
    H[g(t)] = 2 J \sum_{k>0} \psi_k^\dagger \left[ \tau^z (g(t)-\cos k)+\tau^y \sin k  \right] \psi_k, 
\end{equation}
in terms of the fermionic operators $\psi_k^\dagger \equiv (\tilde{c}_k^\dagger, \tilde{c}_{-k}^\phdagger)$ in momentum space. Here, $\tau^{x,y,z}$ are Pauli matrices. We choose to work with periodic boundary conditions so that the momentum is a good quantum number and takes the values $k=(2n+1)\pi/N$ with $n=-N/2,\dots,N/2-1$, as discussed, e.g., in Refs.~\cite{Damski14,SupplInf}. Momentum conservation restricts the formation of defects to kink-antikink pairs. Choosing the total number of spins $N$ to be even proves convenient since the number of kink pairs is then restricted to outcomes in the set $\{0,1,2,\dots,N/2\}$. Given Eq.~\eqref{eq:H_TFIM_fermions}, the dynamics of the TFQIM can be reduced to that of an ensemble of non-interacting two-level systems~\cite{Dziarmaga05}.

Consider a quench, in a finite time $\tau_Q$, from the paramagnetic to the ferromagnetic phase
\begin{equation}
\label{eq:gt}
g(t)=g_c\left(1-\frac{t}{\tau_Q}\right),
\end{equation}
where $g(0)=g_c=1$ is the critical value of $g$, and we let $t$ run from $-3\tau_Q$ to $\tau_Q$. We will refer to $\tau_Q$ as the quench time. We choose $g(\tau_Q)=0$ for simplicity since the final Hamiltonian contains only the ferromagnetic term and commutes with the kink-pair number operator $K_N \equiv \frac{1}{4} \sum_{l=1}^{N} \left(1-\sigma^z_l\sigma^z_{l+1}\right)$. This observable counts the number of kink-antikink pairs in a given quantum state and is extensive in the system size $N$~\cite{Dziarmaga05}. The study of its eigenvalue statistics provided the basis of previous studies exploring universality beyond the KZM~\cite{delCampo2018Universal,Cui20,Bando20,Mayo21,King22}. We define an intensive kink-pair density operator 
\begin{equation}
    \hat{\rho}_N \equiv \frac{K_N}{N}
    = \frac{1}{4N} \sum_{l=1}^{N}  \left(1-\sigma^z_l\sigma^z_{l+1} \right).
\end{equation}
The density of kink pairs, upon completion of the quench in Eq.~\eqref{eq:gt}, is given by the expectation value $\rho_{\rm KZM} = \av{\hat{\rho}_N}$ at the final time $\tau_Q$. It exhibits a power-law scaling in the slow driving limit, i.e., to leading order in a $1/\tau_Q$ expansion~\cite{Polkovnikov05,ZDZ05,Damski05,Dziarmaga05,Polkovnikov11}
\begin{equation}
    \rho_{\rm KZM} = \av{\hat{\rho}_N}=  \frac{1}{4\pi}\sqrt{\frac{\hbar}{2J\tau_Q}},
\end{equation}
in agreement with the celebrated, universal KZM power-law scaling $\rho_{\rm KZM}\propto\tau_Q^{-\frac{\nu}{1+z\nu}}$ for the critical exponents $\nu=z=1$ of the TFQIM~\cite{DZ14}.

In any quantum state other than an eigenstate of $H(g=0)$, the density operator $\hat{\rho}_N$ will exhibit fluctuations of either classical or quantum nature. The probability distribution function $P(\rho_N)$, characterizing the eigenvalue statistics of the kink-pair density operator, reads 
\begin{equation}
    P(\rho_N) = \Av{ \delta(\hat{\rho}_N -\rho_N) },
\end{equation}
where $\rho_N$ is the random variable associated with the kink-pair-number operator  $\hat{\rho}_N$.
 We aim at uncovering via LDT the universality of large fluctuations of $P(\rho_N)$ away from the mean, which the conventional KZM predicts.


{\it Large deviations theory beyond the KZM in the TFQIM.}
The central object in LDT is the scaled cumulant generating function, associated with a random variable $\rho_N$, depending on a large parameter $N$,
\begin{equation}
    \label{eq:def_lambda}
    \lambda(\theta) = \lim_{N\rightarrow \infty} \frac{1}{N}\ln \Av{ e^{N\theta \hat{\rho}_N} }.
\end{equation}
The G\"artner-Ellis theorem states that when $\lambda(\theta)$ exists for all real values of $\theta$, then the random variable $\rho_N$ satisfies the large deviations principle~\cite{Ellis06,Touchette09}
\begin{equation}
    P \big(\rho_N\in[\rho,\rho+d\rho] \big)\approx e^{-NI(\rho)}d\rho,
\label{LDTPeq}
\end{equation}
with the rate function $I(\rho)$ given by the Legendre-Fenchel transform
\begin{equation}
    I(\rho)= \sup_{\theta\in\mathbb{R}} \, \big[ \theta \rho-\lambda(\theta) \big].
\end{equation}
Deviations from the mean value are thus exponentially suppressed by the rate function $I(\rho)$ weighted with the system size, and the random variable concentrates around the mean in the thermodynamic limit. 

Let us consider the application of the G\"artner-Ellis theorem to the distribution of kink pairs generated across a quantum phase transition in the TFQIM. In this case, the defect density is a non-negative quantity. As a result, $I(\rho)$ is divergent for $\rho<0$, and we focus on the case with $\rho \geq 0$. We note that in Fourier space, the operator associated with the density of kink pairs at the end of the quench is
\begin{equation}
    \hat{\rho}_N=\frac{1}{N}\sum_{k>0} \gamma_k^\dagger(\tau_Q) \gamma_k^\phdagger(\tau_Q),
\end{equation}
where $\gamma_k^\phdagger(\tau_Q)$ and $\gamma_k^\dagger(\tau_Q)$ are the fermionic Bogoliubov operators at the end of the quench, and the sum is restricted to $k>0$ since the number of kink pairs equals the number of right-moving kinks. Further, for free fermions (with periodic boundary conditions), the time-dependent density matrix $\varrho(t)$ retains the tensor product structure during unitary time evolution, i.e.,  $\varrho(t) = \bigotimes_k \varrho_k(t)$. As a result, the moment-generating function admits the explicit form
\begin{align}
    \Av{ e^{N\theta \hat{\rho}_N} } &= \prod_{k>0}\Tr\left[\varrho_k(\tau_Q) e^{\theta  \gamma_k^\dagger(\tau_Q) \gamma_k^\phdagger(\tau_Q)}\right] \nonumber\\
    \label{eq:prod_k}
    &= \prod_{k>0} \left[1+ \left(e^{\theta }-1 \right)p_k \right],
\end{align}
where $p_k =\av{\gamma_k^\dagger(\tau_Q) \gamma_k^\phdagger(\tau_Q)} \in[0,1]$ represents the probability that the mode $k$ is excited at the end of the protocol. This is the moment-generating function of a Poisson binomial distribution associated with the sum of $N/2$ independent random Bernoulli variables, each of which has probability $p_k$ for the occupation number to be $1$, corresponding to the formation of a kink-antikink pair, and probability $(1-p_k)$ for the occupation number to be $0$, corresponding to no defect formation~\cite{delCampo2018Universal}. In addition, the value of $p_k$ can be estimated according to the Landau-Zener (LZ) approximation~\cite{Dziarmaga05}, $p_k = \av{ \gamma_k^\dagger(\tau_Q) \gamma_k ^\phdagger(\tau_Q)} \approx \exp(-2\pi J\tau_Qk^2/\hbar)$ near $k=0$, dictating an exponential decay with increasing quench time and a Gaussian decay as a function of the wavenumber. This behavior dictates the KZM scaling in a quantum phase transition~\cite{Dziarmaga05,Damski05,DamskiZurek06}. The explicit computation of the scaled cumulant generating function, according to Eqs.~\eqref{eq:def_lambda} and \eqref{eq:prod_k}, in the limit $N \to \infty$, yields
\begin{equation}
    \lambda(\theta) = \int_0^\pi \frac{dk}{2\pi} \, \ln \left[1+ \left(e^{\theta} - 1 \right) p_k \right],
    \label{eq:lambda_TFIM}
\end{equation}
which is a convergent integral.  For slow quenches, using a power-series expansion in $1/\tau_Q$ to leading order, or equivalently  extending the upper limit of the integral in Eq.~\eqref{eq:lambda_TFIM} to infinity, one finds
\begin{equation}
    \label{SCGFeq}
    \lambda(\theta) = -\rho_{\rm KZM} \mathrm{Li}_{3/2} \left(1-e^{\theta} \right),
\end{equation}
in terms of the polylogarithm function ${\rm Li}_q(x)=\sum_{s=1}^\infty x^s/s^q$.
The exact expression in the slow-quench limit, Eq.~\eqref{SCGFeq}, shows that $\lambda(\theta)$ is differentiable for all values of $\theta$, ensuring the applicability of the G\"artner-Ellis theorem. We verify that for $\theta=0$, $\lambda(0)=0$, consistently with its definition. Further, for $\theta<0$, $\lambda(\theta)$ quickly approaches the constant value $\lambda(-\infty)=-\rho_{\rm KZM}\zeta(3/2)$, where $\zeta$ is the Riemann zeta function. Indeed, $\lambda(\theta)$ is approximately constant for $\theta<0$ and is a monotonic function of $\theta$.

\begin{figure}[t]
    \centering
    \includegraphics[width=\columnwidth]{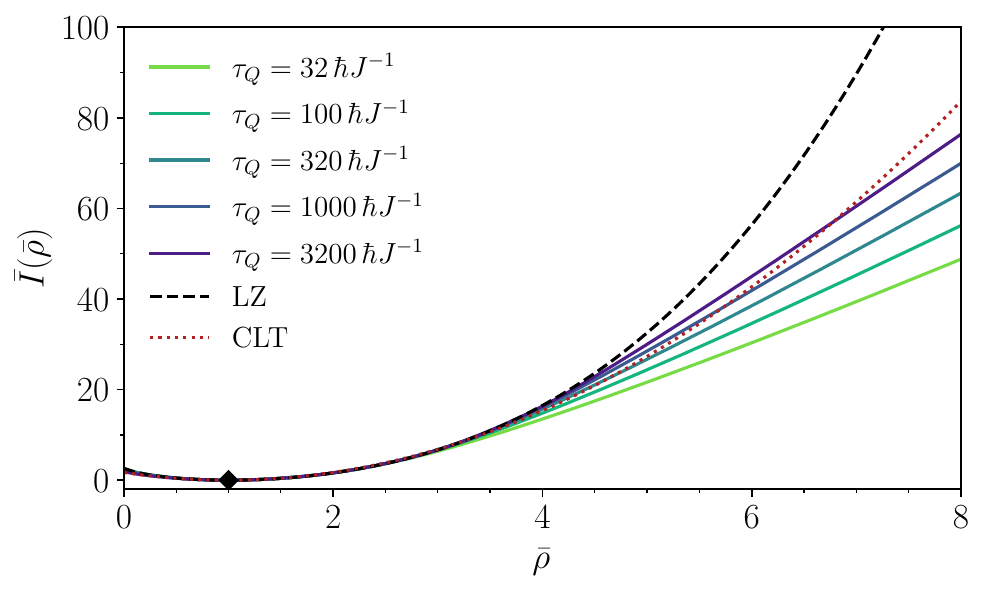}
    \caption{Comparison of the scaled rate function $\bar{I}(\bar{\rho})= I(\rho)/\rho_\mathrm{KZM}$ derived analytically with the numerically exact computation for a finite $\tau_Q$ and $N$. A TFQI chain, initialized in its ground state, is driven by varying $g(t)$ from $g(-3\tau_Q) = 4g_c$ to time $\tau_Q$, when $g(\tau_Q)=0$. The cumulant generating function $\lambda(\theta)$ is computed in the final nonequilibrium state using Eq.~\eqref{eq:def_lambda} for finite $N$, from which the scaled rate function $\bar{I}$ is found with a Legendre-Fenchel transform. As the quench time increases, the agreement between numerics and the analytical solution based on the LZ approximation improves visibly, while the agreement with the central limit theorem (CLT) prediction, obtained by matching the first and second cumulants, does not. The value at the origin $\bar{I}(0) = \zeta(3/2)$ follows from Eq.~\eqref{eq:I_universal}, while the minimum $\bar{I}=0$ is attained at $\bar{\rho}=1$ (diamond). Finite-size analysis reveals the convergence of the numerically-evaluated  rate function $\bar{I}$ to the thermodynamic limit for $N=1000$, which is used in this figure.}
    \label{fig:I}
\end{figure}

We define a dimensionless density of defects $\bar{\rho} \equiv \rho/\rho_{\rm KZM}$ in terms of which
\begin{equation}
    \label{eq:I_universal}
    I(\rho) =\rho_\mathrm{KZM} \sup_{\theta\in\mathbb{R}} \left[\theta \bar{\rho}+{\rm Li}_{3/2} \left(1-e^{\theta}\right)\right].
\end{equation}   
As a result, the  rate function~\eqref{eq:I_universal} is universal in the sense that $\bar{I}(\bar{\rho})= I(\rho)/\rho_\mathrm{KZM}$ varies only with $\bar{\rho}$  and is independent of the quench time $\tau_Q$. This is the central result of our work, which we elaborate and generalize in what follows.
Taking the derivative with respect to $\theta$, one finds at the supremum $\theta^{*}$
\begin{equation}
    \label{eq:thetastar_universal}
    \bar{\rho} = -\frac{e^{\theta^{*}}}{e^{\theta^{*}}-1}\mathrm{Li}_{1/2} \left(1-e^{\theta^{*}}\right).
\end{equation}
The function $\theta^{*}(\bar{\rho})$ and the rate function scaled  by the KZM density $\bar{I}(\bar{\rho})$ are found numerically. The rate function is shown in Fig.~\ref{fig:I}. As the decay of the probability density function $P(\rho)$ is dictated by the rate function according to Eq.~\eqref{LDTPeq}, the minimum of $\bar{I}(\bar{\rho})$ at $\bar{\rho}=1$ is associated with the most likely value of $\hat{\rho}_N$, which equals the mean value $\rho_{\rm KZM}$ predicted by the KZM. Thus, LDT guarantees that the KZM prediction holds with maximum probability. Figure~\ref{fig:I} shows also that only the very large deviations of the defect density $\rho$ are sensible to the finite value of $\tau_Q$. In particular, the larger the quench time $\tau_Q$, the closer the scaled rate function $\bar{I}$ to the universal analytical prediction obtained using the LZ approximation.

{\it Concentration inequalities.} 
Let us tackle the problem of large deviations from a complementary angle using concentration inequalities~\cite{Vershynin18}. To bound large deviations, we make use of the Chernoff bound, which reads $P(\rho_N>\rho) \leq \av{e^{\theta \hat{\rho}_N} } e^{-\theta \rho}$, for all $\theta>0$. The characteristic function can be written as
\begin{align}
    \Av{ e^{\theta \hat{\rho}_N} } &= \exp\left\{ \frac{N}{2\pi} \int_0^\pi dk \, \ln \left[1+(e^{\theta }-1)p_k \right]\right \}\\
    &\approx \exp \left[-N \rho_{\rm KZM}{\rm Li}_{3/2}\left(1-e^{\theta}\right) \right].\nonumber
\end{align}
We thus find from the Chernoff bound that
\begin{equation}
    P(\rho_N>\rho) \leq \exp \left\{-\rho_{\rm KZM} [\theta\bar{\rho}+{\rm Li}_{3/2}(1-e^{\theta})] \right\}
\end{equation}
for all $\theta >0$. Tightening the above inequality by taking the supremum of the exponent, the right tail of the distribution is bounded as 
\begin{equation}
    P(\rho_N>\rho) \leq \exp[-N I(\rho)],
    \label{lefttail}
\end{equation}
with $I(\rho)$ given by Eq.~\eqref{eq:I_universal}. Likewise, the left tail is bound by the same term, $P(\rho_N<\rho) \leq \exp[-N I(\rho)]$. The logarithm of the two-sided Chernoff bound is the rate function. The above results establish the nature of large deviations of kink-antikink pairs formed across the quantum phase transition between the paramagnetic and the ferromagnetic phase of the TFQIM.  These results are generalizable to the family of quasi-free fermion models in which the density of defects is given by the density of quasiparticles. In addition,  the universal form of the scaled cumulant generating function and the rate function in the slow quench limit also hold when fast decaying long-range interactions are considered~\cite{SupplInf}, further confirming their universality under fast-decaying long-range deformations. We next turn our attention to an arbitrary continuous phase transition described by the KZM.

{\it LDT beyond KZM: General scenario.}
Consider a scenario of spontaneous symmetry breaking leading to the generation of point-like defects in $d$ spatial dimensions. The KZM exploits the equilibrium scaling relations for the correlation length $\xi$ and the relaxation time $\tau$, i.e., 
\begin{equation}
    \xi=\frac{\xi_0}{|\varepsilon|^{\nu}}, \qquad
    \tau=\frac{\tau_0}{|\varepsilon|^{z\nu}},
\end{equation}
where $\nu$ and $z$ are critical exponents and $\xi_0$ and $\tau_0$ are microscopic constants. The dimensionless variable $\varepsilon=(g_c-g)/g_c$ quantifies the distance to the critical point $g_c$, and vanishes at the phase transition. Linearizing the driving protocol in the neighborhood of $g_c$ as $g(t)= g_c(1-t/\tau_Q)$, one identifies the quench time $\tau_Q$. The KZM sets the nonequilibrium correlation length to be $\hat{\xi}=\xi_0 (\tau_{Q}/\tau_0)^{\frac{\nu}{1+z\nu}}$~\cite{Zurek85,Zurek96c}. During the phase transition, the system is partitioned into protodomains of average volume $\hat{\xi}^d$. A defect may be generated with an empirical probability $p$ at the merging point between adjacent domains. For point-like defects, the number of events for defect formation is estimated as $\mathcal{N}= V_d/(f\hat{\xi}^d)$, where $V_d$ is the volume in $d$ spatial dimensions and $f$ a fudge factor of order one~\cite{LZ97,Yates98,GomezRuiz20}. As a result,the number of events scales as $\mathcal{N}=[V_d/(f \xi_0^d)]\left(\tau_0/\tau_Q\right)^{\frac{d\nu}{1+z\nu}}$. Assume defect formation events at different locations to be described by independent and identically distributed discrete random variables $X_i$ with $i=1,\dots,\mathcal{N}$~\cite{GomezRuiz20,Mayo21,delcampo21,GomezRuiz22}, where the outcome $X_i=+1$ corresponds to the formation of a topological defect, and $X_i=0$ to no defect formation. The defect number distribution takes the binomial form $P(n)=\binom{\mathcal{N}}{n}p^n(1-p)^{\mathcal{N}-n}$.  Numerical studies support this prediction in $d=1,2$ for varying $\tau_Q$~\cite{GomezRuiz20,Mayo21,delcampo21,GomezRuiz22}. Accordingly, the average number of topological defects is given by $\rho_{\rm KZM}=p \mathcal{N}/V_d \propto \tau_Q^{-\frac{d\nu}{1+z \nu}}$. The defect density, an intensive random variable, is given by $\rho_{\mathcal{N}}=\sum_{i=1}^{\mathcal{N}}X_i/V_d$. We are interested in estimating the probability distribution of $S_{\mathcal{N}}=\sum_{i=1}^{\mathcal{N}}X_i$ when $\mathcal{N}$ is large. Using the Stirling approximation, one finds 
\begin{equation}
    P(S_{\mathcal{N}} = r\mathcal{N}) = \frac{1}{\sqrt{2\pi r(1-r)\mathcal{N}}}e^{-V_d \rho_{\rm KZM}D_{\rm KL}(r\|p)},
\end{equation}
where $D_{\rm KL}(r\|p)=r\ln(r/p)+(1-r)\ln[(1-r)/(1-p)]$ is the Kullback-Leibler distance between two Bernoulli distributions with success probabilities $r$ and $p$. It satisfies $D_{\rm KL}(r\|p) \geq 0$, with the equality holding when $r$ equals $p$, which is the most probable value. Neglecting the prefactor, we thus find that for large $\mathcal{N}$, fluctuations of the defect number away from the mean are suppressed exponentially with increasing $\mathcal{N}$, i.e., $P(S_{\mathcal{N}}=r\mathcal{N})\approx \exp[-V_d\rho_{\rm KZM}D_{\rm KL}(r\|p)]$. In this sense, the defect number distribution concentrates at the KZM prediction in the thermodynamic limit when $V_d$ and $\mathcal{N}$ diverge. Indeed, in the spirit of LDT, we identify the rate function
\begin{eqnarray}
I(r)=\rho_{\rm KZM}D_{\rm KL}(r\|p)\label{eq:rate-classical},
\end{eqnarray}
generalizing the findings for the TFQIM to arbitrary continuous phase transitions.  
Similarly, $I(r)$ dictates the universal suppression of deviations away from the KZM prediction. For example, the right tail of the distribution is bounded as 
\begin{eqnarray}
P(S_{\mathcal{N}}\geq r\mathcal{N})&\leq&\exp\left[-V_d\, I(r)\right]\\
&=&\exp\left[-\frac{pV_d}{f\xi_0^d}\left(\frac{\tau_0}{\tau_Q}\right)^{\frac{d\nu}{1+z\nu}}\!\!D_{\rm KL}(r\|p)\right].\nonumber
\end{eqnarray}
These results are fully consistent with LDT.
Using the dimensionless density of defect $\bar{\rho}\equiv \rho_{\mathcal{N}}/\rho_{\rm KZM}=\frac{1}{p \mathcal N}\sum_{i=1}^{\mathcal{N}}X_i$, 
according to Sanov's theorem in LDT~\cite{Touchette09} 
$P(p\bar{\rho}=r) = e^{-\mathcal{N}D_{\rm KL}(r\|p)}$.
Thus, $P(\bar{\rho}) = \exp[-V_d I(\bar{\rho})]$ 
where $I(\bar{\rho})\equiv \rho_{\rm KZM} D_{\rm KL}(\bar{\rho}p\|p)/p $. The tails of the distribution read then $P(\bar{\rho}\ge\sigma) \leq \exp[-V_d\, I(\sigma)]$ as in Eq. (\ref{lefttail}).

{\it Discussion.} The rate function governs the nature of large deviations away from the mean, according to LDT. Using the exact solution of the critical dynamics in the TFQIM as a test bed, we have explored the nature of large deviations in the number of topological defects generated across a quantum phase transition driven in finite time, and showed that the rate function is proportional to the KZM density of kink pairs. The rate function exhibits a universal power-law scaling with the quench time in which the phase transition is crossed.  We have further generalized these findings to account for the dynamics of arbitrary continuous phase transitions described by the KZM. We have thus proved the KZM, showing that the defect density concentrates at the KZM prediction in the thermodynamic limit, and provided a framework to characterize universal deviations in current experiments with moderate system sizes. Our results are of broad interest in nonequilibrium quantum and classical statistical mechanics, connecting large deviations with the breakdown of adiabatic dynamics, and should find broad applications in quantum simulation, quantum annealing, ultracold atom physics, and the study of critical phenomena.

{\it Acknowledgements.} AdC  thanks SISSA for its hospitality during the early stages of this work.  We thank Federico Roccati for his feedback on the manuscript. This project was funded within the QuantERA II Programme that has received funding from the European Union’s Horizon 2020 research and innovation programme under Grant Agreement No.\ 16434093. AG acknowledges financial support from the
PNRR MUR project PE0000023-NQST. For open access and in fulfillment of the obligations arising from the grant agreement, the authors have applied a Creative Commons Attribution 4.0 International (CC BY 4.0) license to any Author Accepted Manuscript version arising from this submission.
\bibliography{LTDKZM}	

\newpage
\onecolumngrid

\begin{center}
\textbf{\large Supplemental Material}
\end{center}

\section{Fermionic base for the Transverse-Field Quantum Ising Model}

To fix the notation,  we briefly review how the transverse-field quantum Ising model (TFQIM) can be mapped onto a set of non-interacting two-level systems~\cite{Dziarmaga10}. Let us start from the chain Hamiltonian
\begin{equation}
    H[g(t)]  =-J \sum_{l=1}^N \left[g(t)\sigma_l^x + \sigma_l^z \sigma_{l+1}^z\right].
\end{equation}
First, by applying on each site the unitary gate
\begin{equation}
    U = U_\mathrm{Hadamard} \sigma^z
    = \frac{1}{\sqrt{2}}
	\begin{pmatrix}
		1	&-1 \\
		1	&1
	\end{pmatrix},
\end{equation}
$H$ can be brought into the equivalent form
\begin{equation}
    \label{app:eq:H_TFIM_Hadamard}
    H[g(t)]  = J \sum_{l=1}^N \left[g(t)\sigma_l^z - \sigma_l^x \sigma_{l+1}^x\right].
\end{equation}
Second, we apply a Jordan-Wigner transformation and we introduce the fermionic operators
\begin{equation}
	c_l^\dagger := e^{\rmi \pi \Sigma_l} \sigma_l^+, \qquad
	c_l := e^{-\rmi\pi \Sigma_l} \sigma_l^-, \qquad
    \sigma_l^z = 2 c_l^\dagger c_l^\phdagger -1, \label{eq:JordanWigner}
\end{equation}
where $\Sigma_l$ is the string
\begin{equation}
	\Sigma_l := \sum_{j=1}^{l-1} c_j^\dagger c_j^\phdagger.
\end{equation}
The Hamiltonian~\eqref{app:eq:H_TFIM_Hadamard}  in terms of the fermionic operators reads
\begin{equation}
	H[g(t)]  = -J \sum_l \left[ c_l^\dagger c_{l+1}^\dagger + c_l^\dagger c_{l+1}^\phdagger - c_l^\phdagger c_{l+1}^\dagger - c_l c_{l+1} - 2 g(t) c_l^\dagger c_l^\phdagger \right] - NJg(t).
\end{equation}

Next, we define the Fourier basis
\begin{equation}
	\tilde{c}_k := \frac{1}{\sqrt{N}} \sum_l e^{-\rmi k l} c_l, \qquad
	\tilde{c}_k^\dagger := \frac{1}{\sqrt{N}} \sum_l e^{\rmi k l} c_l^\dagger,
\end{equation}
where the symmetric definition of the Fourier transform (i.e.,\ using the $1/\sqrt{N}$ prefactor) is essential to have the commutation relations $\{\tilde{c}_k^\phdagger, \tilde{c}_q^\dagger\} = \delta_{kq}$ without spurious factors of $N$. One needs to impose the boundary conditions to fix the values that $k$ can acquire~\cite{Damski14}. We assume periodic boundary conditions in the spin chain, i.e.,  $\vec{\sigma}_{l+N} = \vec{\sigma}_l$. In the fermionic representation, there is a dependence on the number of particles. If it is even, a particle must acquire a negative phase when performing a full circle. This means $P^+ c_{l+N} P^+ = -c_l$, where $P^+$ projects onto the even sector. Similarly, $P^- c_{l+N} P^- = c_l$. We choose to work with an even number of particles, whence
\begin{equation}
	c_{l+N} = -c_l \quad \implies \quad
	\frac{1}{\sqrt{N}} \sum_k e^{\rmi k l} \tilde{c}_k = \frac{1}{\sqrt{N}} \sum_k e^{\rmi k (l+N)} \tilde{c}_k \quad \implies \quad
	e^{\rmi kN} = -1.
\end{equation}
One can verify that the solution is
\begin{equation}
	k =\frac{2\pi}{N}\left( n + \frac{1}{2} \right)  \qquad \text{for } n =-\frac{N}{2}, -\frac{N}{2}+1,\dots, \frac{N}{2}-1.
\end{equation}

At this point, the fermionic Hamiltonian rewritten in momentum space reads
\begin{equation}
	H[g(t)]  = 2J \sum_{k>0} \left[ -{\rm i} \sin k \left( \tilde{c}_k^\dagger \tilde{c}_{-k}^\dagger + \tilde{c}_{-k} \tilde{c}_k \right) + \left(g(t)-\cos k\right) \left( \tilde{c}_k^\dagger \tilde{c}_k^\phdagger - \tilde{c}_{-k}^\phdagger \tilde{c}_{-k}^\dagger \right) \right].
\end{equation}
Alternatively,
\begin{equation}
	H[g(t)]  = 2J \sum_{k>0}
	\begin{pmatrix}
		\tilde{c}_k^\dagger, 	&\tilde{c}_{-k}^\phdagger
	\end{pmatrix}
	\begin{pmatrix}
		g(t)-\cos k	&-{\rm i}\sin k \\
		{\rm i}\sin k		&-g(t)+\cos k
	\end{pmatrix}
	\begin{pmatrix}
		\tilde{c}_k \\
		\tilde{c}_{-k}^\dagger
	\end{pmatrix}
\end{equation}
and, defining
\begin{equation}
	\psi_k^\dagger := 
	\begin{pmatrix}
		\tilde{c}_k^\dagger, 	&\tilde{c}_{-k}^\phdagger
	\end{pmatrix}, \qquad
	\psi_k := 
	\begin{pmatrix}
		\tilde{c}_k \\
		\tilde{c}_{-k}^\dagger
	\end{pmatrix},
\end{equation}
the Hamiltonian acquires the compact form
\begin{equation}
    \label{app:eq:H_TFIM_fermions}
	H[g(t)] = 2 J \sum_{k>0}h_k[g(t)],
\end{equation}
where
\begin{equation}
h_k[g(t)] \equiv \psi_k^\dagger \left[ (g(t)-\cos k)\tau^z +\sin k \, \tau^y \right] \psi_k,
\end{equation}
and the Pauli matrices $\tau^{x,y,z}$ act on the two entries of the $\psi_k$'s. Notice that $\psi_k$ has only positive $k$'s in the subscript. Equation~\eqref{app:eq:H_TFIM_fermions} is exactly the same as Eq.~\eqref{eq:H_TFIM_fermions} in the main text.

\section{Diagonalization of the Transverse-Field Quantum Ising Model}

We want to diagonalize the Hamiltonian~\eqref{eq:H_TFIM_fermions} through a Bogoliubov rotation. Thus, we change the basis for the creation/annihilation operators as
\begin{equation}
	\begin{pmatrix}
		\gamma_k(t) \\
		\gamma_{-k}^\dagger(t)
	\end{pmatrix} =
	\begin{pmatrix}
		\cos [\theta_k(t)/2]	   &-\rmi \sin [\theta_k(t)/2] \\
		-\rmi \sin [\theta_k(t)/2]  &\cos [\theta_k(t)/2]
	\end{pmatrix}
	\begin{pmatrix}
		\tilde{c}_k \\
		\tilde{c}_{-k}^\dagger
	\end{pmatrix}.
\end{equation}
Again, only $k>0$ are used, leaving an explicit minus sign where needed. The value of $\theta_k$ is fixed by the diagonalization of $h_k$, which yields the relation
\begin{equation}
	\cos \theta_k(t) = \frac{g(t)-\cos k}{\sqrt{g^2(t) + 1 - 2g(t)\cos k}}.
\end{equation}
It follows that
\begin{equation}
	H[g(t)] = \sum_{k>0}
	\begin{pmatrix}
		\gamma_k^\dagger(t), 	&\gamma_{-k}(t)
	\end{pmatrix}
	\begin{pmatrix}
		\epsilon_k(t) 	&0 \\
		0			&-\epsilon_k
	\end{pmatrix}
	\begin{pmatrix}
		\gamma_k(t) \\
		\gamma_{-k}^\dagger(t)
	\end{pmatrix},
\end{equation}
with
\begin{equation}
	\epsilon_k(t) = 2 J\sqrt{g^2(t) + 1- 2g(t)\cos k}.
\end{equation}
Using $\{\gamma,\gamma^\dagger\}=1$, one can rewrite the Hamiltonian as
\begin{equation}
	H[g(t)] = \sum_{k} \epsilon_k(t) \left(\gamma^\dagger_k(t) \gamma^\phdagger_k(t) - \frac{1}{2} \right),
\end{equation}
where the sum runs over both positive and negative $k$'s.

\section{Additional plots}

\begin{figure}
    \centering
    \includegraphics[width=\textwidth]{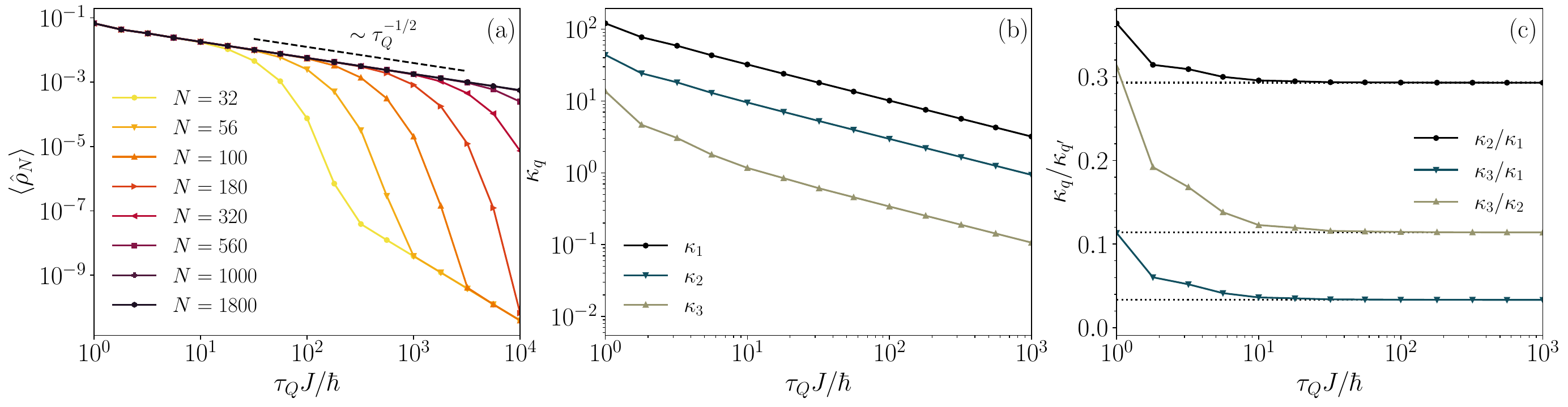}
    \caption{(a) Average kink-pair density produced by the finite-time crossing of the phase transition. One can see the curves collapsing on the single power-law scaling $\av{\hat{\rho}_N} \sim \tau_Q^{-1/2}$ as the quench time is increased. However, at finite $N$ and for quench times too large, the KZM breaks down since the dynamics becomes purely adiabatic.
    (b) Cumulants of the probability distribution $P(\rho)$. As already observed in Refs.~\cite{delCampo2018Universal,Mayo21,Cui20,Bando20,King22}, the cumulants scale with the same power law as a function of the quench time, a hallmark of the underlying Poisson-binomial distribution.
    (c) The ratios between cumulants fastly approach the universal values $\kappa_2 / \kappa_1 = 0.293\dots$, $\kappa_3 / \kappa_1 = 0.0334\dots$, $\kappa_3 / \kappa_2 = 0.114\dots$, shown as dotted lines. To make this figure, the same quench protocol of Fig.~\ref{fig:I} was employed.}
    \label{fig:cumulants}
\end{figure}

We perform additional numerical analysis of the kink statistics in the driven TFQIM, establishing its regime of universality.
In Fig.~\hyperref[fig:cumulants]{\ref{fig:cumulants}(a)} we plot the average density of kink pairs. As predicted by the conventional KZM, the power-law scaling $\av{\hat{\rho}_N} \propto \tau_Q^{-1/2}$ holds if quench times are not too long, i.e., before the onset of adiabaticity. From this figure, it can be checked that the parameter values used to make Fig.~\ref{fig:I} lie well within the KZ scaling region, characterized by the universal power-law scaling with the quench time.

To further establish the regime governed by a universal power-law scaling, in Fig.~\hyperref[fig:cumulants]{\ref{fig:cumulants}(b)--(c)} we compare the scaling of the first three cumulants of the kink-pair probability distribution $P(\rho)$, as a function of the quench time $\tau_Q$. Making use of the fact that the kink-pair distribution is Poisson-Binomial, the explicit expression for the first few cumulants in the TFQIM reads~\cite{delCampo2018Universal,Cui20,Bando20}
\begin{equation}
    \kappa_1 = \sum_{k>0} p_k, \qquad
    \kappa_2 = \sum_{k>0} p_k(1-p_k), \qquad
    \kappa_3 = \sum_{k>0} p_k(1-p_k)(1-2p_k).
\end{equation}
We obtain the excitation probabilities $p_k$  numerically without relying on the LZ approximation. The universal full counting statistics of topological defects  beyond KZM~\cite{delCampo2018Universal,Mayo21,Cui20,Bando20,King22} dictates that all cumulants exhibit a universal power-law scaling with the quench rate, with cumulant ratios being constant. Deviations from this universal behavior are apparent for moderate quench rate satisfying  $J\tau_Q/\hbar<10$, with high-order cumulants being more sensitive than the mean to nonuniversal effects. Panels (b)--(c) of Fig.~\ref{fig:cumulants} identify the regime in which the universal power-law scaling holds, not only for the mean but also for higher-order cumulants. 

\begin{figure}[t]
    \centering
    \includegraphics[width=0.7\textwidth]{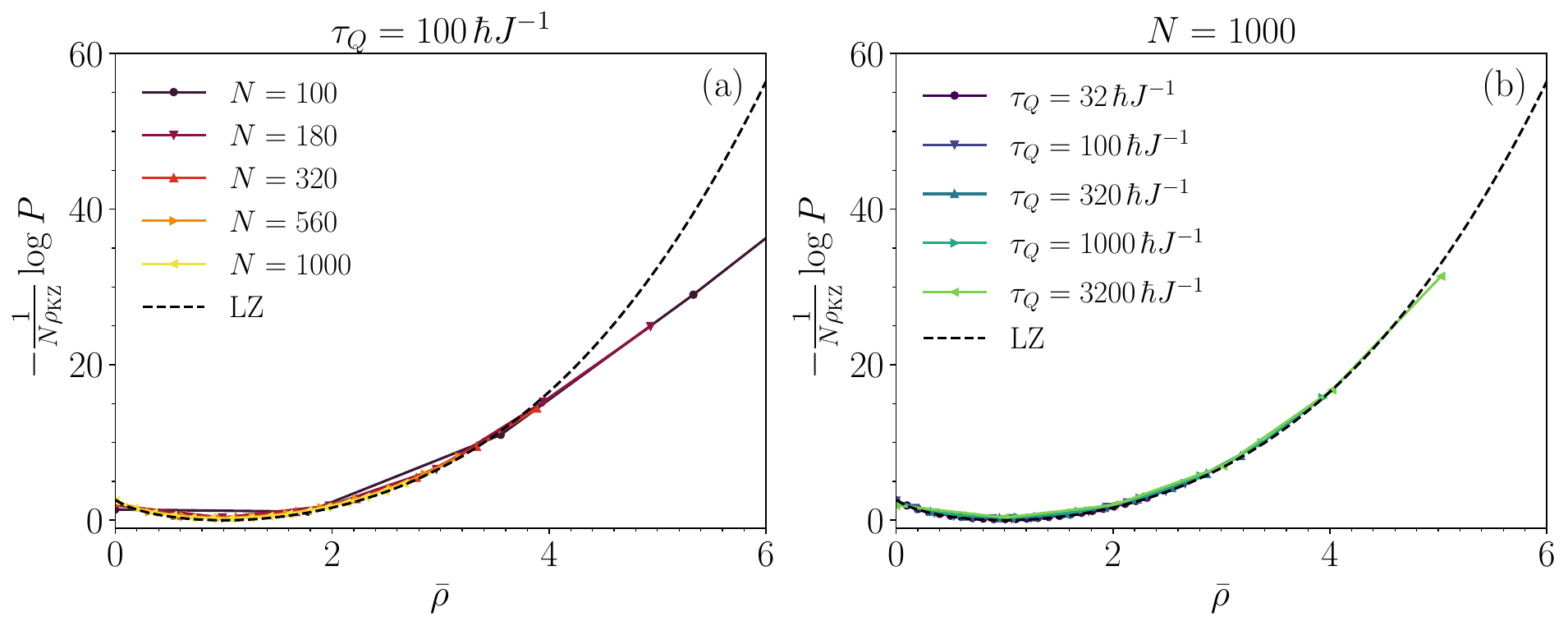}
    \caption{The rate function $\bar{I}$ can be equivalently accessed by computing directly the logarithm of the probability density function. However, even for moderately large $N$, the discreteness of the number of kink pairs results in few data points. To make this figure, the same quench protocol of Fig.~\ref{fig:I} was employed.}
    \label{fig:log_p}
\end{figure}

We complete this supplementary analysis with Fig.~\ref{fig:log_p}, where we explore the convergence of the rate function in LDT from numerically-exact simulations using a finite system size. Specifically, we compared the numerically-evaluated expression $-\frac{1}{N}\ln P(\bar{\rho})$ with the theoretical asymptotic value derived in the main text. Given the finite system size $N$,  the random variable $\hat{\rho}_N$ can assume only the discrete values $1/N, 2/N, \dots, 1/2$. Thus, only a few data points fall into the relevant region of $\bar{\rho}$, when the data is appropriately scaled. By contrast, the computation of $\bar{I}(\bar{\rho})$ makes use of continuous values of $\bar{\rho}$, leading to the better collapse of Fig.~\ref{fig:I}. As a result, the rate function obtained by numerically-exact simulations converges to the thermodynamic limit for moderate system sizes $N=100-1000$. Deviations from the analytical expression derived in the main text simply arise from the fact that the probabilities $p_k$ deviate from the LZ approximation for moderate quench times. This is consistent with the fact that the LZ and the validity of the KZM power-law scaling are restricted to the limit of slow quenches.

\section{Universality of LDT beyond KZM in  systems with fast-decaying long-range interactions}

In this section, we first argue that near the critical region $g=1$, the spectrum of excitations in the long-range deformation of the TFQIM model, i.e., Eq.~\eqref{eq:H-LRalp} is the same as the one in the TFQIM for $\alpha\ge2$. Then we show that the effect of the long-range is to renormalize the quench time by a positive constant. Therefore in the slow quench limit, the universal KZM scaling and the rate function in the LDT are robust against fast-decaying long-range interactions. As such, we expect the rate function Eq.~\eqref{eq:I_universal} in the main text also holds for systems with fast decaying long-range interactions.

\subsection{The long-range Ising and Kitaev model }

To explore the robustness of our predictions in long-range systems, we consider a long-range Hamiltonian   
\begin{eqnarray}
    H_{\text{LR}}[g(t)] & = & -J\sum_{j=1}^{N}\sigma_{j}^{z}\sigma_{j+1}^{z}-Jg(t)\sum_{j=1}^{N}\sigma_{j}^{x}+\frac{J}{2}\sum_{j=1}^{N}\sum_{l=2}^{N-1}\kappa_{l,\alpha}(\sigma_{j}^{z}\sigma_{j+l}^{z}-\sigma_{j}^{y}\sigma_{j+l}^{y})\bigotimes_{n=1}^{l-1}\sigma_{j+n}^{x},
    \label{eq:H-LRalp}
\end{eqnarray}
where 
\begin{equation}
    \kappa_{l,{\alpha}}=
    \begin{cases}
        \frac{1}{l^{\alpha}}        & 1\le l\le N/2\\
        \frac{1}{(N-l)^{\alpha}}    & \text{otherwise},
    \end{cases}
\end{equation}
As for the short-range case, we use periodic boundary conditions $\sigma_{1}^{a}=\sigma_{N+1}^{a}$. We note that $H_{\text{LR}}$, like its short-ranged counterpart, possesses a $\mathbb{Z}_{2}$ symmetry, i.e.,
\begin{equation}
    \left(\bigotimes_{i}\sigma_{i}^{x}\right)H_{\text{LR}}\left(\bigotimes_{i}\sigma_{i}^{x}\right)=H_{\text{LR}}.
\end{equation}
As before, upon using the Hadamard gate to rotate the axes 
\begin{equation}
    \sigma_{j}^{x}\to-\sigma_{j}^{z},\qquad
    \sigma_{j}^{y}\to\sigma_{j}^{y},\qquad
    \sigma_{j}^{z}\to\sigma_{j}^{x},
\end{equation}
one obtains
\begin{equation}
    H_{\text{LR}}[g(t)]=-J\sum_{j=1}^{N}\sigma_{j}^{x}\sigma_{j+1}^{x}+Jg(t)\sum_{j=1}^{N}\sigma_{j}^{z}+\frac{J}{2}\sum_{j=1}^{N}\sum_{l=2}^{N-1}\kappa_{l,\alpha}(\sigma_{j}^{x}\sigma_{j+l}^{x}-\sigma_{j}^{y}\sigma_{j+l}^{y})\bigotimes_{n=1}^{l-1}(-\sigma_{j+n}^{z}).
\end{equation}
Restricting to the even fermionic parity sector, we perform the Jordan-Wigner transformation defined in Eq.~\eqref{eq:JordanWigner} and obtain
\begin{equation}
    H_{\text{LRK}}[g(t)]=-J\sum_{j=1}^{N}(c_{j}^{\dagger}c_{j+1}+c_{j+1}^{\dagger}c_{j})+2Jg(t)\sum_{j=1}^{N}\bigg(c_{j}^{\dagger}c_{j}-\frac{1}{2}\bigg)-J\sum_{j=1}^{N-1}\sum_{l=1}^{N-j}\kappa_{l,\alpha}(c_{j}c_{j+l}+c_{j+l}^{\dagger}c_{j}^{\dagger}).
    \label{eq:H-LRK}
\end{equation}
As explained above, the boundary conditions at the level of the fermionic Hamiltonian become anti-periodic: $c_{j}=-c_{N+j}$. Eq.~\eqref{eq:H-LRK} is the one-dimensional long-range Kitaev (LRK) chain with power-law decaying superconducting pairing~\citep{Vodola2014}. After a Fourier transform, Eq.~\eqref{eq:H-LRK} can be rewritten as
\begin{equation}
    H[g(t)]=2J\sum_{k>0}h_{k}[g(t)],\label{eq:H-fourier}
\end{equation}
where 
\begin{align}
    h_{k}[g(t)] & \equiv\psi_{k}^{\dagger}\left[(g(t)-\cos k)\tau^{z}+f_{\alpha k}\tau^{y}\right]\psi_{k},\nonumber \\
    & =[g(t)-\cos k](\tilde{c}_{k}^{\dagger}\tilde{c}_{k}-\tilde{c}_{-k}\tilde{c}_{-k}^{\dagger})-\mathrm{i}f_{\alpha k}(\tilde{c}_{k}\tilde{c}_{-k}-\tilde{c}_{-k}^{\dagger}\tilde{c}_{k}^{\dagger}),
    \label{eq:Hk-Fourier}
\end{align}
and the function $f_{\alpha k}$ is defined as
\begin{equation}
    f_{\alpha k}\equiv\frac{1}{2}\sum_{l=1}^{N-1}\kappa_{l,\alpha}\sin(kl),
    \label{eq:f-def}
\end{equation}
with the quasi-momentum taking values $k=\pm \frac{1}{2}\frac{2\pi}{N},\,\pm \frac{3}{2}\frac{2\pi}{N}\cdots,\,\pm\frac{2\pi}{N}(\frac{N}{2}-\frac{1}{2})$. In the case of $\alpha=\infty$, it holds that $f_{\infty k}=\sin k$, and the model reduces to the short-range Kitaev chain, i.e., the TFQIM after Jordan-Wigner-transforming back.

Equation~\eqref{eq:H-fourier} can be further diagonalized through the Bogoliubov transformation in the form
\begin{equation}
    H[g(t)]=\sum_{k>0}\epsilon_{k}(t)\psi_k^{\dagger}U_{k}^{\dagger}(t)\tau^{z}U_{k}(t)\psi_k,
\end{equation}
where
\begin{equation}
    \epsilon_{k}(t)=2J\sqrt{[f_{\alpha k}]^{2}+[g(t)-\cos k]^{2}},
    \label{eq:calEps}
\end{equation}
and
\begin{equation}
    U_{k}(t)\equiv
    \begin{pmatrix}
        \cos\left[\theta_{k}(t)/2\right] & -\text{i}\sin\left[\theta_{k}(t)/2\right]\\
    -\text{i}\sin\left[\theta_{k}(t)/2\right] & \cos\left[\theta_{k}(t)/2\right]
    \end{pmatrix},
\end{equation}
with
\begin{equation}
    \sin\theta_{k}(t)=\frac{f_{\alpha k}}{\sqrt{[f_{\alpha k}]^{2}+[g(t)-\cos k]^{2}}},\quad\cos\theta_{k}(t)=\frac{[g(t)-\cos k]}{\sqrt{[f_{\alpha k}]^{2}+[g(t)-\cos k]^{2}}}.
\end{equation}
Denoting 
\begin{equation}
    \begin{pmatrix}
        \gamma_{k}(t)\\
        \gamma_{-k}^{\dagger}(t)
    \end{pmatrix}
    \equiv U_{k}(t)
    \begin{pmatrix}
        \tilde{c}_{k}\\
        \tilde{c}_{-k}^{\dagger}
    \end{pmatrix},
\end{equation}
the Hamiltonian can be rewritten as 
\begin{equation}
    H[g(t)]=\sum_{k}\epsilon_{k}(t)\left[\gamma_{k}^{\dagger}(t)\gamma_{k}(t)-\frac{1}{2}\right],
    \label{eq:H-diagonal}
\end{equation}
where the summation is over all the modes, both positive and negative.

\subsection{Properties of $f_{\alpha k}$}

Let us list a few properties of $f_{\alpha k}$. First of all, $f_{\alpha k}$ is odd in $k$, i.e.,
\begin{equation}
    f_{\alpha k}=-f_{\alpha,-k}.
    \label{eq:f-odd}
\end{equation}
Note that in the thermodynamic limit, the quasi-momentum $k\in[-\pi,\,\pi]$. While $\lim_{k\to \pm\pi}f_{\alpha k}=0$, one should note that $\lim_{k\to0}f_{\alpha}(k)$ does not necessarily vanish. In general, if $\kappa_{l,\alpha}$ decays slowly as $l$ increases, $f_{\alpha k}$ may become singular at $k=0$, as shown in Ref.~\citep{yang2022superheisenberg}. In fact, one can express $f_{\alpha k}$ in terms of the polylogarithm function
\begin{equation}
    f_{\alpha k}=- {\rm i}\left[\text{Li}_{\alpha}(e^{{\rm i}k})-\text{Li}_{\alpha}(e^{-{\rm i}k})\right].
\end{equation}
Using the expansion of polylogarithm function, one can show that $\alpha\notin\mathbb{N}^{+}$, $f_{\alpha k}\propto 1/k^{1-\alpha}$~\citep{Vodola2014}, which is adopted by Ref.~\citep{dutta2017probing} to discuss KZM in the long-range Kitaev chain. We note that this scaling behavior can be alternatively proved using the Euler-Maclaurin formula for any $\alpha\ge0$ without the constraint $\alpha\notin\mathbb{N}^{+}$. Thus, in view of Eq.~\eqref{eq:f-odd}, $f_{\alpha k}$ allows the following expansion near $k=0$,
\begin{equation}
    f_{\alpha k}=\frac{\xi_{0}}{k^{1-\alpha}}+\xi_{1}k+O(k^{3}).
    \label{eq:f-Taylor}
\end{equation}

\subsection{LDT for the kink density with fast-decaying long-range interactions}

The phase diagram of the long-range Kitaev chain is very rich~\citep{Vodola2014,pezz`e2017multipartite}, including in particular topological phase transitions. With the spectrum Eq.~\eqref{eq:calEps}, for $\alpha>1$ one can easily verify that the gap closes only when $g=\pm1$, with the gapless mode being $k=0$. Furthermore, according to Eq.~\eqref{eq:f-Taylor}, if one only considers $\alpha\ge2$, then
\begin{equation}
    f_{\alpha k}= k +\mathcal{O}(k^3).
\end{equation}
This is the same behaviour of the spectrum of the TFQIM near $k=0$, where $f_{\alpha k}=\sin k$. Clearly, for both $-1<g<1$ and $g>1$, one can continuously deform $\alpha$ from $2$ to $\infty$ without closing the gap. As we will show below, the kink density statistics and the large deviations theory are the same in the slow quench limit. We consider the ramp in the main text, i.e.,
\begin{equation}
    g(t)=g_{c}\left(1-\frac{t}{\tau_{Q}}\right),
\end{equation}
starting at an initial time before $-\tau_{Q}$, outside the freeze-out region~\cite{DZ14}, and ending at $\tau_{Q}$, whence $g(\tau_Q)=0$. We  investigate the statistics of kink density and work with the adiabatic basis at $t=-\infty$, 
\begin{gather}
    \ket{0}, \qquad \ket{k,-k}=c_{k}^{\dagger}c_{-k}^{\dagger}\ket{0} = \gamma_{k}^{\dagger}(-\infty)\gamma_{-k}(-\infty)\ket{0}, \\
    \ket{k,0}=c_{k}^{\dagger}\ket{0}=\gamma_{k}^{\dagger}(-\infty)\ket{0}, \qquad
    \ket{0,-k}=c_{-k}^{\dagger}\ket{0}=\gamma_{-k}(-\infty)\ket{0}.
\end{gather}
For each mode Hamiltonian  $Jh_{k}[g(t)]$ preserves the fermionic parity, and excitations  remain in the subspace spanned by $\ket{0},\ket{k,-k}$. Projecting Eq.~\eqref{eq:Hk-Fourier} onto this subspace, one obtains the effective Hamiltonian 
\begin{equation}
    Jh_{k}[g(t)]=J
    \begin{pmatrix}
        [g(t)-\cos k] & -{\mathrm{i}}f_{\alpha k}\\
        {\mathrm{i}}f_{\alpha k} & -[g(t)-\cos k]
    \end{pmatrix}.
    \label{hamiltonian:two-level}
\end{equation}
The excitation generated during the process can be characterized by the Landau-Zener formula, provided $|k|<\pi/2$:
\begin{equation}
    p_{k}=\Av{\gamma_{k}^{\dagger}(\tau_Q)\gamma_{k}(\tau_Q)} \approx \exp\bigg(-\frac{2\pi J\tau_{Q}}{\hbar}f_{\alpha}^{2}(k)\bigg),
\end{equation}
where the average is over the state of the system at the final time $t=\tau_Q$. The  scaled cumulant generating function becomes resembles that in the TFQIM
\begin{equation}
    \lambda(\theta)=\frac{1}{2\pi}\int_{0}^{\pi}dk\ln\bigg[1+(e^{\theta}-1)\exp\bigg(-\frac{2\pi J\tau_{Q}}{\hbar}f_{\alpha}^{2}(k)\bigg)\bigg],
\end{equation}
where we have extended the integral up to $\pi$, an approximation with exponential accuracy. When $\alpha\ge2$, according to Eq.~\eqref{eq:f-Taylor}, 
\begin{equation}
    f_{\alpha k}=\xi k+O(k^{3}).
\end{equation}
In the slow quench limit where $\tau_{Q}\to\infty$, one can perform a Gaussian approximation, leading to 
\begin{equation}
    p_{k}= \Av{\gamma_{k}^{\dagger}(\tau_Q)\gamma_{k}(\tau_Q)}=\exp\bigg(\frac{-2\pi J\tau_{Q}\xi^{2}k^{2}}{\hbar}\bigg),
\end{equation}
and thus
\begin{equation}
    \lambda(\theta)=\frac{1}{2\pi}\int_{0}^{\pi} dk\ln\bigg[1+(e^{\theta}-1)\exp\bigg(\frac{-2\pi J\tau_{Q}\xi^{2}k^{2}}{\hbar}\bigg)\bigg].
\end{equation}
Comparing the results for TFQIM, we see that the effect of the fast decaying long-range interactions is to renormalize the quench time by a positive constant, keeping the structure of the scaled cumulant generating function otherwise unchanged.

\end{document}